# Compact Broadband Light Source Based on Noise-Like Pulses

FANGLIN CHEN,[1,2] XIAHUI TANG,[1] MING TANG,[1] AND LUMING ZHAO[1,2,*]

[1]*School of Optical and Electronic Information and Wuhan National Laboratory for Optoelectronics, Optics Valley Laboratory, Huazhong University of Science and Technology, Wuhan, 430074, China*
[2]*Shenzhen Huazhong University of Science and Technology Research Institute, Shenzhen, China*
*lmzhao@ieee.org

**Abstract:** We report on broadband generation based on noise-like pulse (NLP) fiber lasers at 1.55 μm and 1.06 μm, respectively. The 1.55 μm laser system can generate a broadband spectrum with a 20 dB bandwidth of up to 205 nm, while the 1.06 μm one can achieve a 20 dB bandwidth of 341 nm after amplification and spectral broadening. Simulation results reproduce experimental details and highlight the role of nonlinear effects in achieving broad spectral outputs, underscoring the suitability of NLPs for advanced applications.

## 1. Introduction

Fiber lasers, known for their exceptional performance, are extensively utilized across a wide range of applications, including industrial manufacturing[1], medical treatments [2], and scientific research [3]. Additionally, fiber lasers provide a crucial platform for investigating nonlinear dynamics, as the generating pulses exhibit a wide range of dynamic behaviors. Through precise control of dispersion and nonlinearity, and by configuring various cavity structures and mode-locking techniques, fiber lasers can produce diverse soliton states, including dissipative soliton resonance, soliton explosions, and harmonic mode-locking. Among these, noise-like pulses (NLPs) have garnered particular attention due to their unique pulse structure. NLPs feature a broader averaged spectrum and higher pulse energy, making them highly promising for applications in areas such as laser processing [4], optical sensing [5, 6], and low-coherence spectral interferometry [7]. In the time domain, NLPs manifest as pulse envelopes on the picosecond to nanosecond scale, with envelope shapes typically rectangular or H-shaped. Within these envelopes are numerous femtosecond sub-pulses, whose peak power and pulse width vary randomly. Despite the chaotic behavior of the sub-pulses, the overall envelope remains stable, circulating within the cavity. Due to the instability of the internal pulse structure, NLPs exhibit low coherence, typically being reflected by a broad pedestal and sharp peak of the autocorrelation trace. The limited acquisition rate of spectrum analyzer results in the NLP spectra observed in experiments being averaged over multiple measurements, yielding smooth, broad spectrum with a full width at half maximum (FWHM) extending to tens or even hundreds of nanometers. NLPs are primarily generated by soliton pulse splitting under high pump power. The numerous femtosecond sub-pulses generated interact within the cavity, potentially leading to extreme phenomena such as rogue waves [8]. These sub-pulses undergo strong nonlinear interactions within the cavity, influenced by effects such as self-phase modulation, four-wave mixing, and stimulated Raman scattering, which collectively broaden the spectrum. Additionally, due to the presence of many ultrashort pulses within the NLP envelope, the total pulse energy is higher. By using double-clad or large-mode-area doped fibers, the single-pulse energy of NLPs can exceed 300 nJ [9]. In thulium-doped fiber lasers, single-pulse energies exceeding 10 μJ have been achieved through a triple-stage external amplification process [10]. NLPs can be generated in both the normal and anomalous dispersion regimes of fiber lasers and have been reported in various mode-locking and cavity configurations, such as nonlinear optical loop mirrors (NOLM) [11, 12], nonlinear amplifying loop mirrors (NALM) [13-15], and semiconductor saturable absorber mirrors (SESAM) [16]. Compared to other supercontinuum generation schemes, fiber lasers provide greater flexibility and compactness in design. Moreover, the use of an all-polarization-maintaining structure can enhance system stability. The high energy and broad spectral characteristics of NLPs significantly reduce the demands on amplification and spectral broadening stages, making the NLP fiber laser a highly promising source for supercontinuum generation. The broad spectrum of NLP allows for the generation of narrowband spectral slices that can be effectively used in wavelength-division multiplexed (WDM) fiber optic communication [17]. After intensity modulation and subsequent signal amplification, such as with an erbium-doped fiber amplifier (EDFA), these spectral slices can significantly enhance data transmission capacity and efficiency in optical networks.

The NLP scheme enables the direct generation of a broad and flat spectrum in fiber lasers. Additionally, an NLP source can serve as a seed, producing supercontinuum through external amplification and spectral broadening. While NLPs can be generated in conventional fiber laser cavities, precise dispersion and nonlinearity management are essential for effectively broadening the spectral width of NLP pulses. Introducing dispersion-shifted fibers (DSF) to the cavity can increase its length, thereby facilitating pulse energy lump sum and further enhancing the spectral broadening process [18]. More importantly, extending the cavity length increases the duration of nonlinear interactions, thereby enhancing spectral broadening through mechanisms such as self-phase modulation and stimulated Raman scattering. Experimental studies on direct dispersion management have demonstrated that the broadest spectral bandwidth is achieved when the dispersion in a Ytterbium-doped fiber (YDF) ring laser cavity approaches zero [19]. By introducing highly nonlinear fibers after the erbium-doped fiber (EDF) for nonlinear management, Zhao et al. achieved a spectral bandwidth exceeding 200 nm in an erbium-doped fiber laser [20]. Compared to ultrashort pulses, NLPs offer significant advantages as seed sources for supercontinuum generation. Their high pulse energy facilitates enhanced nonlinear interactions, while their broad spectral characteristics contribute to the flatness of the resulting supercontinuum. By combining an NLP seed source with highly nonlinear fiber, Xia et al. demonstrated a supercontinuum spectrum with a 20 dB bandwidth exceeding 500 nm. The NLP, with an initial bandwidth of 60 nm, was generated by an erbium-doped fiber laser and subsequently broadened using 110 m of highly nonlinear fiber (HNLF) [21]. In another instance, using an EDF laser mode-locked by a NOLM, an NLP was generated. By combining this NLP source with an EDFA and 5 m of HNLF, a supercontinuum spectrum with a 20 dB bandwidth of 1000 nm was achieved [22].

In this paper, we report on the development of 1.55 μm and 1.06 μm NLP lasers. The 1.55 μm NLP laser directly generates a flat, broadband spectrum with a 20 dB bandwidth of 205 nm, while the 1.06 μm NLP laser, being followed by external amplification and spectral broadening, achieves a 20 dB bandwidth of 341 nm. These results highlight the significant advantages of NLPs as broadband light sources.

## 2. NLP generation in EDF lasers

The configuration of the erbium-doped noise-like broadband fiber laser is shown in Fig. 1. The laser employs nonlinear polarization rotation (NPR) to achieve mode-locking. To ensure the high-gain condition necessary for generating NLPs, two 976 nm pump sources are used for bidirectional pumping. The cavity includes a 2-meter-long erbium-doped fiber (EDF 80 from OFS), followed by a 9-meter-long HNLF (NL-1550-NEG from YOFC). To facilitate integration, two 0.3-meter-long ultra-high numerical aperture fiber (UHNA, UHNA7 from Coherent) segments are spliced at both ends of the HNLF as bridging fibers. Testing with a 1550 nm light source revealed that the inclusion of the HNLF and UHNA fibers introduced approximately 4.4 dB of additional loss to the cavity. Furthermore, to increase the single-pulse energy of the generated NLPs, a 62-meter-long single-mode fiber is incorporated, which lowers the pulse repetition rate and promotes nonlinear interaction accumulated. The total cavity length is approximately 106 meters, and a 5% output coupler is employed to preserve the majority of the pulse energy inside the cavity.

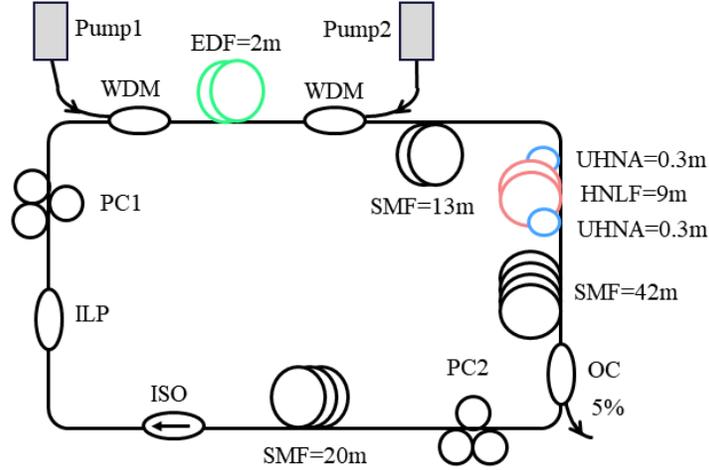

Fig. 1. Schematic of the 1.55μm fiber laser. PC: Polarization controller; OC: output coupler; ISO: Isolator; ILP: In-line polarizer

The forward pump power is set to 835mW, and the backward pump power is set to 350 mW. By carefully adjusting the polarization controller (PC), we successfully obtained NLPs, as illustrated in Fig. 2. The temporal waveform of the pulse, as shown in Fig. 2(a), was recorded by using a 1 GHz high-speed oscilloscope (Tektronix, MDO34) paired with a 5-GHz photodetector (Thorlabs, DET08CFC/M). The measured output power is 4.07 mW (Thorlabs, PM100D). The pulse train consists of NLP envelopes, with a repetition interval of approximately 530 ns, corresponding to the cavity length. Figure 2(b) provides further details of the NLPs, showing that each envelope contains several sub-pulses spanning a time scale of approximately 200 ns. The sub-pulses exhibit random intensity and positioning, resulting in the decoherence of the NLPs. The output spectrum was analyzed using an optical spectrum analyzer (OSA, Yokogawa AQ6370D). Figure 2(c) presents the average spectrum of the output pulses. Due to the limited wavelength range of the OSA, long-wavelength component beyond 1700 nm can not be detected, which limits the capture of the entire NLP spectrum. Despite this limitation, the spectrum within the detectable range demonstrates remarkable broad bandwidth, with a 3 dB bandwidth of 82.56 nm. The spectral lines are exceptionally flat across this width, reflecting the superior spectral characteristics of the laser. Notably, the 20 dB bandwidth exceeds 205.37 nm, underscoring the broad and stable spectrum achieved.

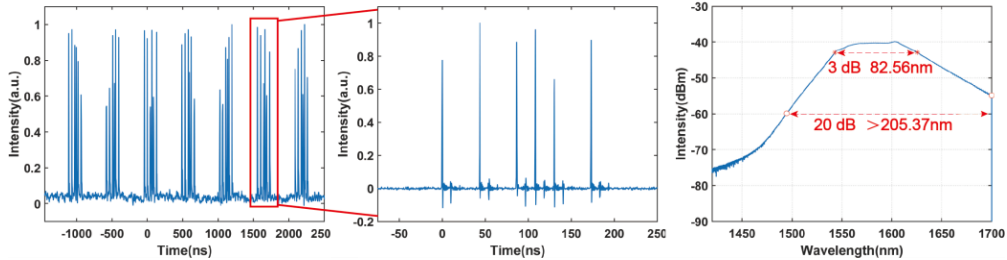

Fig. 2. Experimental results. (a) Oscilloscope trace. (b) temporal zoom-in of NLP wave packet in the time domain (c) Output spectrum.

To elucidate the physical mechanism and influencing factors of NLP generation, we performed numerical simulations based on the cubic complexed Ginzburg–Landau equation (CGLE). The CGLE is a well-established model for describing the evolution of ultrashort pulses in dissipative systems, such as fiber lasers. The equation is formulated as follows:

$$\frac{\partial U}{\partial t} = -i\frac{\beta_2}{2}\frac{\partial^2 U}{\partial t^2} + \frac{\beta_3}{6}\frac{\partial^3 U}{\partial t^3} + \frac{g}{2}U + \frac{g}{2\Omega^2}\frac{\partial^2 U}{\partial t^2} + i\gamma\left(1 + i\tau_{shock}\frac{\partial}{\partial t}\right) \quad (1)$$

$$U(z,T)\int_{-\infty}^{\infty} R(T')|U(z,T-T')|dT'$$

where U are slowly varying envelopes propagating along the fiber. To simplify the simulation, we consider the dispersion parameters of the EDF and those of the single mode fiber are same, where $\beta_2 = -23\,ps^2/km$, $\beta_3 = 0.13\,ps^3/km$, $\gamma = 3(W\cdot km)^{-1}$. $g$ is the gain provided by the EDF and $g = 0$ in the single mode fiber. In the EDF, $g$ is defined as:

$$g = g_0 \exp\left(-\frac{1}{E_{sat}}\int |u|^2\,dt\right) \quad (2)$$

where $g_0$ is the small signal gain and $E_{sat} = 1.2\,nJ$ is the saturable energy. The saturable absorption effect produced by NPR is represented as:

$$T = 1 - \frac{q}{1+|u|^2/I_{sat}} - \alpha \quad (3)$$

where $q$ represents the modulation depth, $\alpha$ denotes the unsaturated loss, and $I_{sat}$ is the saturation intensity. For the EDF, we considered its typical gain spectrum, as shown by the black dashed line in Fig. 3. For the HNLF, we set $\gamma = 10(W\cdot km)^{-1}$ and $\beta_2 = -2.42\,ps^2/km$ according to the vendor's data sheet. The response of the fiber to the pulses includes both the instantaneous response and the delayed Raman response, which are expressed as:

$$R(t) = (1-f_R)\delta(t) + f_R h_R(t) \quad (4)$$

where $h_R = \frac{\tau_1^2 + \tau_2^2}{\tau_1 \tau_2^2}\exp(-t/\tau_2)\sin(t/\tau_1)$, The Raman response function is typically approximated by $h_R$, in which $\tau_1 = 12.2\,fs$, $\tau_2 = 32\,fs$. $f_R$ is the fractional contribution of the delayed Raman response, typically around 0.18 for silica fibers.

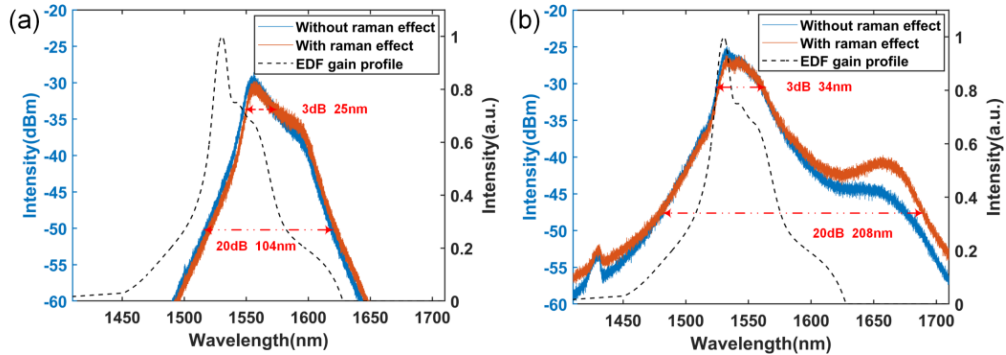

Fig. 3. The simulation results under different cavity parameters.

In the simulation, we neglected the self-steepening effect. The cavity state was adjusted by tuning $g_0$ and $I_{sat}$, resulting in two distinct NLP states. To explore the impact of the Raman effect on the NLP spectrum, the value of $f_R$ in Equation 3 was modified (either $f_R = 0$ or $f_R = 0.18$), while keeping all other parameters constant. First, by $g_0 = 9000 m^{-1}$ setting and $I_{sat} = 100W$, the NLP reached a stable state after several round trips in the cavity. The corresponding spectrum is depicted in Fig. 3(a). At this stage, the spectral peak had shifted from the central wavelength of the EDF emission spectrum. After propagating through the HNLF, the pulse underwent splitting, and the intracavity gain was sufficient to maintain the presence of smaller sub-pulses. The spectral peak, located at 1556 nm, extended to 1590 nm, exhibiting a relatively flat profile with an average spectral intensity fluctuation of approximately 5 dB. The 3 dB bandwidth was measured to be around 25 nm, while the 20 dB bandwidth spanned about 104 nm. Due to the existence of multitude of smaller sub-pulses, the sub-pulse amplitudes were relatively low, and the Raman effect had minimal influence on the overall spectrum.

Upon further adjustment of parameters $g_0 = 12500 m^{-1}$ and $I_{sat} = 110W$, the pulse stabilized, and the corresponding spectrum is shown in Fig. 3(b). At this point, the spectral peak aligned with the central wavelength of the EDF emission spectrum. The 3 dB bandwidth increased to approximately 34 nm, and the 20 dB bandwidth expanded significantly to about 208 nm. The reduction in the number of sub-pulses led to an increase in average amplitude, and the Raman peak at 1650 nm became more pronounced.

## 3. NLP generation in YDF lasers

The NLP laser can function as a broadband seed source for external amplification and spectral broadening. We constructed an YDF laser to generate NLP pulses. Mode-locking is achieved by using NPR technique again. The laser is forward-pumped by using a 976 nm pump source. The YDF (CorActive Yb501) has a length of 65 cm, while components such as the wavelength division multiplexer (WDM), isolator (ISO), inline polarizer (ILP), output coupler (OC), and PCs account for the majority of the total cavity length, which is approximately 14 meters. The OC is selected to deliver 30% output, ensuring adequate energy in the resulting pulses. Externally, a 41 cm section of YDF, pumped with a 976 nm pump source, constitutes the amplification stage. To monitor the pulse properties before and after amplification, 99:1 OCs are positioned at key points. Following amplification, a 20-meter-long photonic crystal fiber (PCF, PC1011-A from YOFC) is used to facilitate spectral broadening.

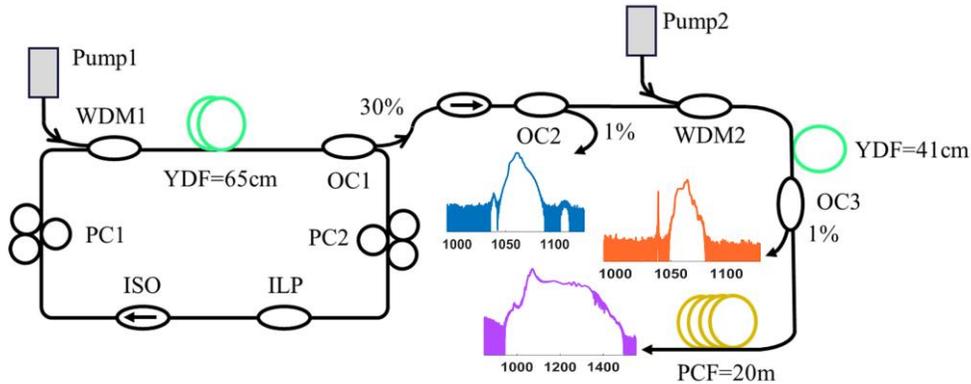

Fig. 4. Schematic of the 1.06μm fiber laser. PC: Polarization controller; OC: output coupler; ILP: In-line polarizer; ISO: Isolator; PCF: Photonic crystal fiber.

The output pulses of the laser are shown in Fig. 4. We set the pump power of the laser to 247 mW. The output power is 45.4 mW, with a repetition rate of 14 MHz, resulting in a pulse energy of 3.24 nJ. The central wavelength of the pulses is located at 1060 nm, and due to the

YDF gain spectrum, there is a side lobe at 1037 nm. The Raman effect shifts some of the pulse energy to 1110 nm. The 3 dB spectral bandwidth of the pulses is 5.62 nm, while the 20 dB bandwidth reaches 30.07 nm. The pump power of the external amplification stage is 370 mW. After passing through the YDF, significant gain is achieved at 1038 nm, resulting in a sharp spectral peak. The amplified pulse power reaches 284.6 mW, with a 3 dB spectral bandwidth of 4.69 nm and a 20 dB bandwidth of 31.14 nm. After propagating through the PCF, the pulses experience effective spectral broadening, with the 20 dB bandwidth expanding to 341 nm. The peak wavelength shifts to 1071 nm, red-shifting by approximately 10 nm compared to the input pulse. The spectral peak introduced by the YDF at 1038 nm remains visible in the broadened spectrum.

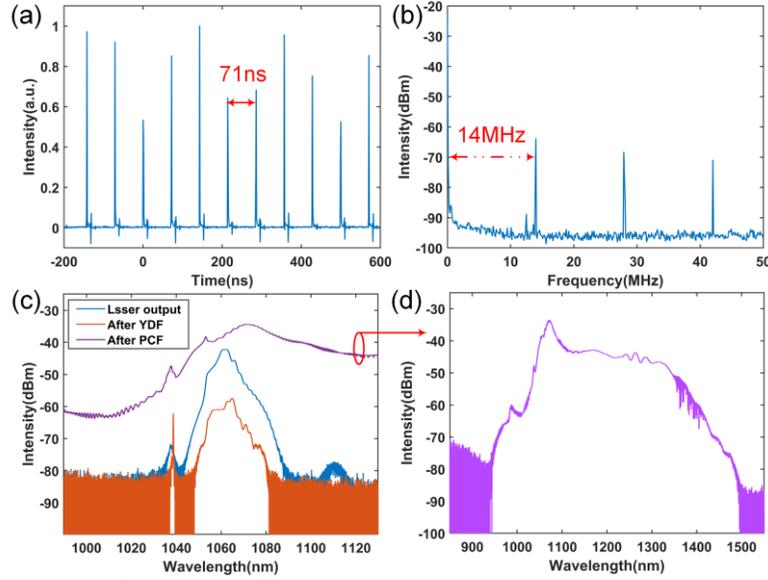

Fig. 5. Experimental results. (a) Oscilloscope trace. (b) RF spectrum. (c) Output spectrum. (d) The wide-range spectrum after PCF.

We qualitatively simulated the spectral broadening process in the PCF. Due to the presence of several sub-pulses within the NLP envelope, we set the peak power of each incident sub-pulse to 1600 W with a pulse width of 0.6 ps, and the parameters of the PCF are $\gamma = 18(W \cdot km)^{-1}$, $\beta_2 = -8.68\,ps^2/km$, $\beta_3 = 0.079\,ps^3/km$. The simulation results are shown in Fig 6. Inside the PCF, the pulse spectrum undergoes significant broadening. The incident pulse, with a central wavelength of 1060 nm, experiences sideband extension reaching up to 1500 nm within the PCF.

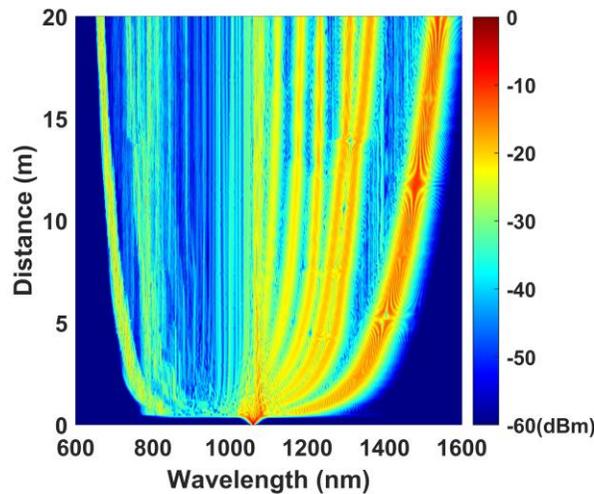

Fig. 6. The simulation of the spectral broadening process in the PCF.

## 4. Conclusion

We have developed and characterized NLP fiber lasers at 1.55 μm and 1.06 μm, respectively, demonstrating their efficacy as broadband light sources. The 1.55 μm NLP laser directly generates a flat spectrum with a 20 dB bandwidth of 205 nm, while the 1.06 μm laser, after external amplification and propagation through a segment of PCF, achieves a 20 dB bandwidth of 341 nm. Both experimental and simulation results expose the role of nonlinearity management in achieving efficient spectral broadening. The demonstrated broad and stable spectra, along with high pulse energies, underline the potential of NLP lasers in supercontinuum generation. This work establishes NLP lasers as a viable and compact solution for broadband light generation, with further optimization offering potential improvements in spectral performance.

## Acknowledgment

This work is supported by the National Key Research & Development Program of China (2023YFB2805101); Hubei Provincial Science and Technology Plan Project (2022EHB001); Fundamental Research Funds for the Central Universities (HUST2020kfyXJJS007); Science and Technology Planning Project of Shenzhen Municipality (JCYJ20220818103214029); Shenzhen Science and Technology Program (JCYJ20220530160607016).

**Disclosures**

The authors declare no conflicts of interest.

**Data availability.** Data underlying the results presented in this paper are not publicly available at this time but may be obtained from the authors upon reasonable request.